\documentclass[nofootinbib,floatfix,preprint,superscriptaddress,a4paper]{revtex4-1}

\usepackage{amsmath,ulem}
\usepackage{amssymb}
\usepackage{graphicx}
\usepackage{color}
\usepackage{subfigure}
\usepackage{url}
\usepackage[T1]{fontenc} 
\usepackage{slashed}

\usepackage{ulem} 

\newcommand{\Br}{\text{Br}}

\newcommand{\AddrBonn}{%
Bethe Center for Theoretical Physics \& Physikalisches Institut der Universit\"at Bonn, \\
Nu{\ss}allee 12, 53115 Bonn, Germany
}

\newcommand{\AddrOrsay}{%
Laboratoire de Physique Th\'eorique, CNRS -- UMR 8627, 
Universit\'e de Paris-Sud 11\\ F-91405 Orsay Cedex, France}

\preprint{BONN-TH-2012-07,LPT-12-37}

\begin{document}

\title{New bounds on trilinear $R$-parity violation from lepton flavor violating observables}

\author{H. K. Dreiner}\email{dreiner@th.physik.uni-bonn.de}

\author{K. Nickel}\email{nickel@th.physik.uni-bonn.de}

\author{F. Staub}\email{fnstaub@th.physik.uni-bonn.de}
\affiliation{\AddrBonn}

\author{A. Vicente} \email{avelino.vicente@th.u-psud.fr}
\affiliation{\AddrOrsay}

\begin{abstract}
Many extensions of the leptonic sector of the Minimal Supersymmetric
Standard Model (MSSM) are known, most of them leading to observable
flavor violating effects.  It has been recently shown that the 1-loop
contributions to lepton flavor violating three-body decays $l_i \to 3
l_j$  involving the $Z^0$ boson may be dominant, that is,
much more important than the usual photonic penguins. Other processes
like $\mu$-$e$ conversion in nuclei and flavor violating $\tau$ decays
into mesons are also enhanced by the same effect. This is for instance
also the case in the MSSM with trilinear
$R$-parity violation. The aim of this work is to derive new bounds on
the relevant combinations of $R$-parity violating couplings
and to compare them with previous results in the literature.
For heavy supersymmetric spectra the limits are improved by several
orders of magnitude. For completeness, also constraints coming from
flavor violating $Z^0$-decays and tree-level decay channels $l \to l_i
l_j l_k$ are presented for a set of benchmark points.
\end{abstract}
\maketitle

\section{Introduction}

Supersymmetry (SUSY) is one of the most popular extensions of the
Standard Model (SM) \cite{Martin:1997ns,Nilles:1983ge}.  It provides a
technical solution to the famous hierarchy problem
\cite{Gildener:1976ai,Veltman:1980mj,Sakai:1981gr,Witten:1981nf} and
 contains the required ingredients to accommodate new physics
\cite{Ellis:1983ew}.

However, no  experimental evidence of supersymmetry has been
found so far at the Large Hadron Collider (LHC)
\cite{Chatrchyan:2008zzk,Aad:2009wy}.  Direct searches, based mainly
on the existence of missing transverse energy in the final state, have
failed to find a signal that exceeds the SM background
\cite{atlasweb,cmsweb}. This  should encourage the search for
non-minimal supersymmetric scenarios with a departure from the usual
supersymmetric signatures. Therefore, new strategies might be 
necessary, such as those required to look for trilinear 
$R$-parity violation ($R$pV) \cite{Hall:1983id,Allanach:2003eb}.

The non-observation of lepton or baryon number violating processes in
nature sets strong bounds on the trilinear $R$-parity violating
couplings. Furthermore, some SM processes are also affected by the
introduction of these couplings, which allows us to set additional
experimental limits. Many studies in this direction can be found, see
for example \cite{Barbier:2004ez,Kao:2009fg,Dreiner:2010ye}.

The lepton flavor violating (LFV) decay $l_i \to 3 l_j$, $i\not
=j$, is a well-known process in supersymmetry. However, although
detailed computations exist in the literature
\cite{Hisano:1995cp,Arganda:2005ji}, some of its properties have been
missed until very recently. The dominance of the photon mediation
diagrams, only affected by Higgs mediation in the large $\tan
\beta$ regime \cite{Babu:2002et}, has been part of the common lore for many
years. This led to the simple relation
\begin{equation}
\Br(l_i \to 3 l_j) \simeq \frac{\alpha}{3 \pi} 
\left[\log\left(\frac{m^2_{l_i}}{m^2_{l_j}}\right) - \frac{11}{4} \right] 
\Br(l_i \to l_j \gamma) \, ,
\end{equation}
which implies $\Br(l_i \to 3 l_j) < \Br(l_i \to l_j \gamma)$. This is
in fact true in the minimal supersymmetric standard model (MSSM)
with lepton flavor violation. Contrary to this, it was recently
pointed out that the $Z^0$-penguin, usually neglected or regarded as a
subleading contribution, can induce a huge enhancement of the signal
in extended models and lead to $\Br(l_i \to 3 l_j) > \Br(l_i \to l_j
\gamma)$ \cite{Hirsch:2012ax}. This implies that some LFV studies need
to be revisited in order to take into account the constraining power
of $l_i \to 3 l_j$.

One of the extended scenarios where the $Z^0$-penguin enhancement is
found is trilinear $R$-parity violation. The additional lepton number
violating interactions, not present in the MSSM, induce a large 1-loop
$\Br(l_i \to 3 l_j)$. This increase has been unnoticed in the existing
literature \cite{Choudhury:1996ia,deGouvea:2000cf}. Furthermore, the
same $Z^0$-penguins will also dominate the amplitudes for $\mu-e$
conversion in nuclei and $\tau \to l_j P^0$ decays (where $P^0$ is a
pseudoscalar meson). We will use these observables
to set new bounds on the combinations of trilinear couplings
involved. Finally, for the sake of completeness, we will also cover
the 1-loop decays $Z^0 \to l_i l_j$ and the tree-level decays $l_i \to
3 l_j$ and $l_i \to l_j l_k l_k$ and refer to Ref.~\cite{Dreiner:2006gu}
for an exhaustive collection of bounds coming from tree-level decays 
involving mesons.

\section{Lepton flavor violating observables in $R$-parity violating SUSY}

In this section we discuss how the flavor violating decays $l_i \to 3
l_j$, $l_i \to l_j l_k l_k$, $Z^0 \to l_j l_k$ as well as $\mu-e$
conversion in nuclei and $\tau \to l_i P^0$ decays are induced in
trilinear $R$-parity violating SUSY. Although the focus of this work
is the impact of the $Z^0$-penguin on
the 1-loop induced $l_i \to 3 l_j$ decays and $\mu-e$ conversion
in nuclei, we also study the loop induced decay $Z^0 \to l_j l_k$. In
addition, the decays at tree-level are given for completeness in the
appendix. 

\subsection{Lepton flavor violating three-body decays: $l_i \to 3 l_j$}
 We start our discussion with the leptonic three-body decay
$l_i \to 3 l_j$, since this process gives a clear understanding of
the impact of the $Z^0$ penguin. The total width of the 1-loop induced $l_i \to 3 l_j$ decay contains
contributions from the photon penguin, the Higgs penguin, the
$Z^0$-penguin and box diagrams. For instance, the amplitudes for the important
photon and $Z^0$ penguins can be written as
\begin{align}
\label{eq:Ampitude1}
T_{\gamma-{\rm penguin}} =& \bar{u}_i(p_1)\left[q^2 \gamma_{\mu} 
(A_1^L P_L + A_1^R P_R) + i m_{l_j} \sigma_{\mu \nu} q^{\nu} \left( 
A_2^L P_L + A_2^R P_R \right) \right] u_j(p) \nonumber \\
\times& \frac{e^2}{q^2}  \bar{u}_i(p_2) \gamma^ {\mu} v_i(p_3) - (p_1 
\leftrightarrow p_2)  \, , \\
\label{eq:Ampitude2}
T_{Z^0-{\rm penguin}} =& \frac{1}{m_Z^2} \bar{u}_i(p_1) \left[ \gamma_
{\mu} \left( F_L P_L + F_R P_R \right) \right] u_j(p) \nonumber \\
\times& \bar{u}_i(p_2) \left[ \gamma^{\mu} \left( Z_L^{(l)} P_L + Z_R^{(l)} P_R \right) \right] v_i(p_3) - (p_1 \leftrightarrow p_2) \, . 
\end{align}
Here $A_{1,2}^{L,R}$ and $F_{L,R}$ represent the 1-loop form factors
induced by the photon and $Z^0$-boson exchange, respectively, and
$Z_{L,R}^{(l)}$ are the standard $Z^0$-boson couplings to the
leptons. The long expressions for the scalar penguins and boxes can be
parametrized by the operators $B^{I}_{L,R}$ (with $I=1, \dots 4$). The
total width $\Gamma \equiv \Gamma(l_i^- \to l_j^- l_j^- l_j^+)$ is
obtained as~\cite{Hisano:1995cp,Arganda:2005ji}:
\begin{widetext}
\begin{eqnarray} \label{1loop-width}
\Gamma &=& \frac{e^4}{512 \pi^3} m_{l_i}^5 \left[ \left| A_1^L \right|^2 + 
\left| A_1^R \right|^2 - 2 \left( A_1^L A_2^{R \ast} + A_2^L A_1^{R \ast} 
+ h.c. \right) \right. \nonumber \\
&+& \left( \left| A_2^L \right|^2 + \left| A_2^R \right|^2 \right) \left( 
\frac{16}{3} \log{\frac{m_{l_i}}{m_{l_j}}} - \frac{22}{3} \right) \nonumber \\
&+& \frac{1}{6} \left( \left| B_1^L \right|^2 + \left| B_1^R \right|^2 
\right) + \frac{1}{3} \left( \left| \hat{B}_2^L \right|^2 + \left| 
\hat{B}_2^R \right|^2 \right) \nonumber \\
&+& \frac{1}{24} \left( \left| \hat{B}_3^L \right|^2 + \left| \hat{B}_3^R 
\right|^2 \right) + 6 \left( \left| B_4^L \right|^2 + \left| B_4^R \right|^2 
\right) \nonumber \\
&-& \frac{1}{2} \left( \hat{B}_3^L B_4^{L \ast} + \hat{B}_3^R B_4^{R \ast} 
+ h.c. \right) \nonumber \\
&+& \frac{1}{3} \left( A_1^L B_1^{L \ast} + A_1^R B_1^{R \ast} + A_1^L 
\hat{B}_2^{L \ast} + A_1^R \hat{B}_2^{R \ast} + h.c. \right) \nonumber \\
&-& \frac{2}{3} \left( A_2^R B_1^{L \ast} + A_2^L B_1^{R \ast} + A_2^L 
\hat{B}_2^{R \ast} + A_2^R \hat{B}_2^{L \ast} + h.c. \right) \nonumber \\
&+& \frac{1}{3} \left\{ 2 \left( \left| F_{LL} \right|^2 + \left| F_{RR} 
\right|^2 \right) + \left| F_{LR} \right|^2 + \left| F_{RL} \right|^2 
\right. \nonumber \\
&+& \left( B_1^L F_{LL}^{\ast} + B_1^R F_{RR}^{\ast}  + \hat{B}_2^L F_{LR}^
{\ast}  + \hat{B}_2^R F_{RL}^{\ast} + h.c. \right) \nonumber \\
&+& 2 \left( A_1^L F_{LL}^{\ast} + A_1^R F_{RR}^{\ast} + h.c. \right) + 
\left( A_1^L F_{LR}^{\ast} + A_1^R F_{RL}^{\ast} + h.c. \right) \nonumber \\
&-& 4 \left. \left. \left( A_2^R F_{LL}^{\ast} + A_2^L F_{RR}^{\ast} + h.c. 
\right) - 2 \left( A_2^L F_{RL}^{\ast} + A_2^R F_{LR}^{\ast} + h.c. \right) 
\right\} \right] \nonumber \, . \\
\label{decay}
\end{eqnarray}
\end{widetext}
Here, $F_{XY}$ are functions of $F_L$ and $F_R$ and the Higgs and box
contributions are combined into $\hat{B}$. Exact definitions can be
found in \cite{Arganda:2005ji}. We do not repeat them here for the
sake of brevity. Finally, $\text{Br}(l_i \to l_j \gamma)$,
$i\not=j$, is completely determined by the same form factors
$A_2^L$ and $A_2^R$
\begin{equation}
\text{Br}(l_i \to l_j \gamma) = \frac{e^2}{16 \pi} m_{l_i}^5 \left(|A_2^L|^2 + |A_2^R|^2 \right) \, .
\end{equation}

For many years the decay $l_i \to 3 l_j$ has been believed to be
dominated by photon exchange, with large Higgs contributions in the
large $\tan \beta$ regime \cite{Babu:2002et}. This has been recently
challenged in Ref.~\cite{Hirsch:2012ax}, where it was shown that many
simple extensions of the leptonic sector can lead to large
enhancements for the $Z^0$ boson contributions. This may lead to
$Z^0$-penguin dominated scenarios where $\Br(l_i \to 3 l_j) > \Br(l_i
\to l_j \gamma)$. In fact, this can be understood from simple
dimensional arguments. As shown in Eq.~\eqref{1loop-width}, the 
decay width is proportional to $m_{l_i}^5$, so both $A$ and $F$ form
factors must have dimensions of inverse mass squared. 
Thus we only have to determine what is the mass scale for each
case. First, the vanishing mass of the
photon implies that the only mass scale involved in the $A$ form
factors is $m_{SUSY}$. On the other hand, the mass scale of
the $F$ form factor is set by $m_Z$, the $Z^0$ boson mass. Therefore, we conclude that
$A \sim m_{SUSY}^{-2}$ and $F\sim m_Z^{-2}$. This fact can be checked
analytically in the complete expressions given in
Refs.~\cite{Hisano:1995cp,Arganda:2005ji}. With $m_Z^{2} \ll
m_{SUSY}^2$ the $Z^0$ penguin can, in principle, be even more
important than the photonic one.

However, in the case of the MSSM the photonic penguin is found to be
numerically dominant \cite{Arganda:2005ji}. This is caused by a subtle
cancellation among the different $Z^0$ boson diagrams
\cite{Hirsch:2012ax} which strongly suppresses  their
contribution to the amplitude of the process. We note that a similar
behavior was found in Ref.~\cite{Lunghi:1999uk} for the decay $B
\to X_s l^+ l^-$.

\begin{figure}
\centering
\includegraphics[width=0.6\linewidth]{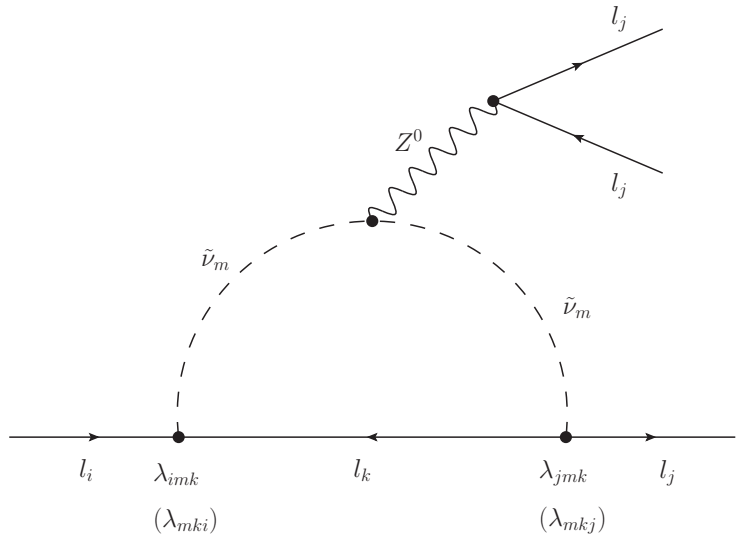}
\caption{1-loop induced $l_i \to 3 l_j$ decays. As shown in brackets, 
there are two possible combinations of $\lambda$ couplings:
$\lambda_{jmk} \lambda_{imk}$ and $\lambda_{mkj}
\lambda_{mki}$. Moreover, we remind the reader that the $\lambda$
couplings are antisymmetric in the first two indices. Similar diagrams
with the $Z^0$ boson line attached to the lepton lines are also
possible.}
\label{1loop-diag}
\end{figure}

However, this cancellation can be easily spoiled by two effects,
either (1) extended particle content, or (2) new interactions in the
lepton sector. Trilinear $R$-parity violation is a simple example of
the second case. The additional interactions of the leptons lead to
new loop diagrams including charged leptons which do not suffer from
the same cancellation as the wino does and induce a large increase in
the $l_i\to 3 l_j$ signal; \textit{cf.} the $Z^0$ mediated
diagrams in Fig.~\ref{1loop-diag}. It is the object of this paper to
study how this increase, together with the current experimental bounds,
constrains the relevant parameter
space. We will also shortly comment on the impact of possible
future improvements on the experimental limit for this observable
 \cite{mueee}.

So far, we have not mentioned decays of the form
$l_i \to l_j l_k l_k$ with different generations of 
leptons in the final states. The reason is that these decays will 
always be less constraining than $l_i \to 3 l_j$ because of combinatorical
factors which lead to $\text{Br}(l_i \to l_j) > \text{Br}(l_i \to l_j l_k l_k)$
\cite{Ilakovac:1994kj}.

\subsection{$\mu - e$ conversion}
Let us now discuss $\mu-e$ conversion in nuclei. This process is also
mediated by photonic, $Z^0$ and Higgs penguins as well as box diagrams
\cite{Arganda:2007jw}.  The $Z^0$ contributions are given by the same 
diagram as shown in Fig.~\ref{1loop-diag} with the two
external leptons attached to the $Z^0$ replaced by quarks. The
conversion rate can be expressed as \cite{Arganda:2007jw}
\begin{align}
\text{Cr}(\mu - e, \text{Nucleus}) = &  \frac{1}{\Gamma_{capt}} \frac{p_e E_e m_\mu^3 G_F^2 \alpha^3 Z_{eff}^4 F_p^2}{8 \pi^2 Z}  \cdot \nonumber \\
 & \hspace{-1cm} \cdot \left( \left| (Z+N)^2(g_{LV}^{(0)} + g_{LS}^{(0)}) + (Z - N) (g_{LV}^{(1)} + g_{LS}^{(1)}) \right|^2 + L \leftrightarrow R \right)
\end{align}
Here, $Z$ and $N$ are the number of protons and neutrons in the
nucleus, $Z_{eff}$ is an effective charge, $F_p$ is the nuclear matrix
element and $\Gamma_{capt}$ denotes the total muon capture rate. The
different contributions $g_{XY}^{(J)}$ ($X=L,R; Y=V,S; J=0,1$) are
functions of the same form factors $A$ and $F$ already introduced in
Eqs.~(\ref{eq:Ampitude1})-(\ref{eq:Ampitude2}) as well as of scalar
penguins and box diagrams. For a detailed discussion we refer to
Ref.~\cite{Arganda:2007jw}.

Similarly, the decays $\tau \to l_i P^0$ get contributions from $Z^0$
mediated diagrams, which lead to the corresponding $F$ form factors,
and from pseudoscalar ($A^0$) mediated diagrams
\cite{Arganda:2008jj}. As for $\mu-e$ conversion in nuclei, one
expects that the $Z^0$-penguins dominate. Furthermore,
it turns out that $\mu-e$ conversion in nuclei and $\tau \to l_i P^0$
are even more constraining than $l_i \to 3 l_j$. 
This is mainly due to the very good existing
experimental limits
\cite{Dohmen:1993mp,Honecker:1996zf,Bertl:2006up}. 
In addition, there
are also very good experimental perspectives, with plans for a sensitivity
for $\mu-e$ conversion rates as low as $10^{-18}-10^{-16}$
\cite{Carey:2008zz,Cui:2009zz}. A detailed comparison
of the importance of the different observables is given in section~\ref{sec:results}.

\subsection{Lepton flavor violating $Z^0$ decays}
As already mentioned, we also present here results for the lepton
flavor violating $Z^0$ decays. These have been discussed in the context of
trilinear $R$-parity violation in
Refs. \cite{Chemtob:1998uq,Yang:2010iq}.  These decays are triggered
by diagrams like the one given in Fig.~\ref{1loop-diag} but without
the two leptons attached to the $Z^0$ boson. The branching ratio can
be expressed as \cite{Bi:2000xp}
\begin{equation}
\text{Br}(Z^0 \to l_i l_j) = \frac{1}{\Gamma_Z}\frac{1}{48 \pi M_Z} \left[2 (|a_1|^2+|a_2|^2) M_Z^2 + \frac{1}{4}(|a_3|^2+|a_4|^2)M_Z^4 \right] \, .
\end{equation}
There is only an explicit suppression by the SUSY scale for the
contributions $a_3$ and $a_4$ but $a_1$ and $a_2$ are dimensionless.
This observable has been discussed in the context of a SUSY $SO(10)$
model in \cite{Bi:2000xp}. Because of this dependence on the different
scales the authors have observed in the considered $SO(10)$ model that
$\text{Br}(Z^0 \to \tau \mu)$ actually increases with increasing
universal scalar mass $m_0$, until it saturates. However, the overall
impact of this observable was found to be rather small because of the
weak experimental limits. We note that a similar behavior was found in
\cite{Yang:2010iq}.

\section{Trilinear $R$-parity violation}
We consider in this work the impact of the $Z^0$ penguins in the MSSM extended 
by the lepton number violating terms \cite{Hall:1983id,Allanach:2003eb}
\begin{equation}
\label{eq:Superpotential}
 W_{\slashed{R}} = \frac{1}{2}  \lambda_{ijk} \hat{L}_i \hat{L}_j \hat{E}^c_k 
 + \frac{1}{2}  \lambda^{'}_{ijk} \hat{L}_i \hat{Q}_j \hat{D}^c_k 
\end{equation}
Bounds for these trilinear couplings have been set so far not only by
using lepton flavor violating decays, but also $\mu - e$ conversion in
nuclei or cosmological observations.  This lead to limits on individual
couplings or specific products of couplings
\cite{Barger:1989rk,Barbier:2004ez,Kao:2009fg,Dreiner:2010ye,Bhattacharyya:1996nj,Allanach:1999ic,Dreiner:2001kc,Dreiner:2006gu}. However, all studies
dealing with Br($l_i \to l_j l_k l_l$) have so far neglected all
contributions but the photonic penguins. Also the bounds from rare
$Z^0$ decays in case of trilinear $R$-parity violation have not been
presented in the literature so far.

Before we discuss the new bounds which arise if one performs the full
calculation including all contributions, we comment shortly on the
bilinear $R$-parity violating term which was skipped in
Eq.~(\ref{eq:Superpotential}). It is well know that the trilinear
couplings will induce also a term $\kappa_i \hat{L}_i \hat{H}_u$
during the RGE evaluation \cite{Banks:1995by,Allanach:2003eb}.  This
term, as well as the corresponding soft-breaking terms $B_{\kappa _i}
H_u\tilde{l}_i$ and $m^2_{H_d l} \tilde{l}^*_i H_d$, lead already at
tree-level to a mixing between standard model and supersymmetric
states.  In addition, they generate small vacuum expectation values
(VEVs) for the sneutrinos\footnote{For these and other aspects of
  bilinear $R$-parity violation and neutrino mass generation see
  Ref.~\cite{Hirsch:2004he} and references therein.}. However, the
values of $\kappa_i$ are restricted by neutrino data and the size of
the additional VEVs by electroweak precision data. Therefore, the
impact of bilinear $R$-parity violation and the related couplings on
the lepton flavor violating decays considered here are in general
sub-dominant and numerically negligible \cite{Hirsch:2012ax}.  The
only exception can be found when a large lepton-chargino mixing, which
can open new tree-level channels, is induced. However, also these
contributions are suppressed by the SUSY scale and might only be
relevant for light spectra \cite{Faessler:2000pn}.

\section{Numerical analysis}
\label{sec:results}
\subsection{Setup}
The numerical analysis has been performed by means of the Fortran
package {\tt SPheno} \cite{Porod:2003um,Porod:2011nf} using the
Mathematica interface provided by {\tt SARAH}
\cite{Staub:2011dp,Staub:2010jh,Staub:2009bi}. 

The Fortran code generated by {\tt SARAH} to calculate $l_i \to 3 l_j$
and $l_i \to l_j \gamma$ is based on the generalization of the
formulas given in Ref.~\cite{Arganda:2005ji}. The routines for $\mu-e$
conversion and $\tau \to l_i P^0$ are based on
Refs.~\cite{Arganda:2007jw} and \cite{Arganda:2008jj}, respectively.
The generic expressions for the rare $Z^0$-decays have been calculated
with {\tt FeynArts} and {\tt FormCalc} \cite{Hahn:2000kx,Hahn:2009bf}
and have been compared with the formul{\ae} of Ref.~\cite{Bi:2000xp}:
while we agree with the vertex correction, our results for the wave
function contributions are smaller by an overall factor of 2. The
output of the {\tt SPheno} code for $\mu-e$ conversion in nuclei,
$\tau \to l_i P^0$ decays and lepton flavor violating $Z^0$ decays will
become a new public feature of {\tt SARAH 3.1.0}.

We want to stress that in case of the three-body decays or $\mu - e$
conversion in nuclei our computation includes not only the photonic
and $Z^0$-penguins but also the contributions from Higgs penguins and
box diagrams.  Finally, {\tt SARAH} writes the routines to calculate
all three-body decays of fermions at tree-level which were used to
obtain the results given in the appendix.

To disentangle the effect of the renormalization group
evaluation we have first calculated the MSSM parameters at the
electroweak scale for three benchmark points given in
Ref.~\cite{AbdusSalam:2011fc}.  These points are called BP1 - BP3 in
the following. In addition, we have included a CMSSM scenario which
leads to sneutrino masses of $\sim$100~GeV (point BP0).  Although this
point leads to a SUSY spectrum already ruled out by LHC searches, it
is presented here to compare the obtained results with the bounds
previously given in the literature. Even BP1 might already be 
borderline, especially as long as $R$-parity violating effects are
small. However, we have included it also here to close the gap between
the old studies in the literature and the points BP2 and BP3 with a
heavy spectrum that satisfy all recent collider bounds. The input
parameters as well as some relevant masses are given in
Table~\ref{tab:points}.  In the table we focus on the relevant
masses for the discussion and skipped those which play a negligible
role in the calculation of the constraints. As expected, the main
result can in general be obtained from the diagram shown in
Fig.~\ref{1loop-diag}. Similar diagrams with neutralinos or charginos
give smaller contributions.

\begin{table}[hbt]
 \begin{tabular}{|c|c|c|c|c|}
 \hline
 \hline
     & BP0 & BP1 & BP2 & BP3 \\
\cite{AbdusSalam:2011fc} &  & 10.1.1 & 10.4.1 & 40.2.5 \\
 \hline
 \multicolumn{5}{|c|}{Input}  \\
 \hline
 $m_0$~[GeV]  &  100 & 125 & 750 & 750 \\
 $M_{1/2}$~[GeV] & 100 & 500 & 350 & 650 \\
 $\tan(\beta)$ & 10 & 10 & 10 & 40 \\
 sign($\mu$) & + &  + & + & + \\
 $A_0$~[GeV] & 0 & 0 & 0 & -500 \\
 \hline
 \multicolumn{5}{|c|}{Masses} \\
 \hline
$\tilde{d}_R,\tilde{s}_R$     &      257.8  &      1017.5 &        1497.0  &         1483.5   \\
$\tilde{d}_L,\tilde{s}_L$     &      261.0  &      1020.9 &        1503.8  &         1532.9  \\
$\tilde{b}_1$         &      240.7  &       975.1  &       1434.2  &         1285.6   \\
$\tilde{b}_2 $        &      269.8  &      1065.9 &        1570.0  &         1364.7  \\
$\tilde{u}_R,\tilde{c}_R$         &      254.7  &     1024.3  &      1509.7    &       1477.8     \\
$\tilde{u}_L,\tilde{c}_L$         &      257.8  &     1063.1  &      1568.1    &       1531.0    \\
$\tilde{t}_1 $        &      190.3  &       812.1  &       1208.8  &          1095.0   \\
$\tilde{t}_2$         &      331.8  &    1021.2   &     1466.1     &          1333.0  \\
$\tilde{e}_R, \tilde{\mu}_R$     &      115.2  &       229.7  &        450.2   &         788.6   \\
$\tilde{e}_L, \tilde{\mu}_L$     &      129.9  &       361.2  &        610.3   &         864.9   \\
$\tilde{\tau}_1$         &      107.8  &       222.1  &        442.5   &         601.8   \\
$\tilde{\tau}_2$         &      134.8  &       362.5  &        611.1   &         801.6   \\
$\tilde{\nu}_e,\tilde{\nu}_\mu$   &      102.0  &       352.2  &        605.7   &         860.6   \\
$\tilde{\nu}_\tau$       &      101.4  &       351.0  &        603.5   &         787.0   \\
\hline\hline
\end{tabular}
\caption{Input parameters as well as relevant SUSY masses for benchmark
points BP0 - BP3. BP1-BP3 correspond to those points of
Ref.~\cite{AbdusSalam:2011fc}, as indicated in the second row of this
table. BP0 is included for comparison with earlier results in the
literature. All masses are given in GeV.}
\label{tab:points}
\end{table}
After the calculation of the MSSM spectrum, we switched on the
different combinations of the $R$pV couplings which can open flavor
violating decay or transition channels and calculated the different 
observables at tree- and 1-loop level. 
The tree-level results are given in the appendix. 

In the determination of the bounds we have used the most recent
experimental upper limits given in Table~\ref{tab:bounds}.
\begin{table}[hbt]
\begin{tabular}{|c|c||c|c||c|c|}
\hline
\hline
$\text{Br}(\mu \to e \gamma)$   & $2.4 \cdot 10^{-12}$ & $\text{Br}(\tau \to e \gamma)$ & $3.3 \cdot 10^{-8}$   & $\text{Br}(\tau \to \mu \gamma)$ & $4.4 \cdot 10^{-8}$ \\
$\text{Br}(\mu \to 3 e)$        & $1.0 \cdot 10^{-12}$ & $\text{Br}(\tau \to 3 \mu)$    &  $ 2.7 \cdot 10^{-8}$ & $\text{Br}(\tau \to 3\mu)$       &$ 2.1 \cdot 10^{-8}$ \\
$\text{Br}(Z^0 \to e \mu)$        & $1.7\cdot 10^{-6}$   & $\text{Br}(Z^0 \to e \tau)$      & $9.8\cdot 10^{-6}$    & $\text{Br}(Z^0 \to \mu \tau)$      &$ 1.2\cdot 10^{-5}$ \\
$\text{Cr}(\mu - e,\text{Pb})$                  &  $4.6 \cdot 10^{-11}$                 & $\text{Cr}(\mu - e ,\text{Ti})$      &$ 6.1 \cdot 10^{-13} $ & $\text{Cr}(\mu - e ,\text{Au})$        &$ 7.0 \cdot 10^{-13}$ \\
$\text{Br}(\tau \to e \pi^0)$ & $ 8.0 \cdot 10^{-8}$ & $\text{Br}(\tau \to e \eta)$ & $ 9.2 \cdot 10^{-8}$  & $\text{Br}(\tau \to e \eta')$  & $ 1.6 \cdot 10^{-7} $ \\ 
$\text{Br}(\tau \to \mu \pi^0)$ & $ 1.1 \cdot 10^{-7}$ & $\text{Br}(\tau \to \mu \eta)$ & $ 6.5 \cdot 10^{-8}$  & $\text{Br}(\tau \to \mu \eta')$  & $ 1.3 \cdot 10^{-7} $ \\
\hline
\hline
\end{tabular} 
\caption{Current experimental upper limits on flavor violating two- and 
three-body decays [$\text{Br}(l_i \to l_j \gamma)$/$\text{Br}(l_i\to3
l_j)$], flavor violating $Z^0$ decays [$\text{Br}(Z^0\to l_il_j)$],
$\mu-e$ conversion rate [$\text{Cr}(\mu-e,X)$] and semi-leptonic,
flavor violating $\tau$ decays ($\tau \to l_i P^0$)
\cite{Dohmen:1993mp,Honecker:1996zf,Bertl:2006up,Adam:2011ch,Nakamura:2010zzi}.}
\label{tab:bounds}
\end{table}

For the 1-loop induced decays, the limits would not be improved if we
also took into account observables with two different generations of
leptons in the final state. This is due to the fact that $\tau^- \to
e^+ \mu^- \mu^-$ and $\tau^- \to \mu^+ e^- e^-$ would only be
triggered by box diagrams which are in general suppressed with respect
to the penguins.  In addition, the branching ratios for decays like
$\tau^- \to e^- \mu^+ \mu^-$ will always be smaller than those for a
single flavor final state. The reason for this can be found in the
relative factors of the $Z^0$ and photon contributions in the
corresponding partial widths. They always lead to $\text{Br}( l_i \to
3 l_j) > \text{Br}(l_i \to l_j l_j l_k)$ ($j \neq k)$, see
Ref.~\cite{Ilakovac:1994kj}.

\subsection{Results for 1-loop induced observables}
The focus in this section is on combinations of $\lambda$ and $\lambda'$ which do not open
flavor violating tree-level decay channels for the leptons if there is not any other source
of lepton flavor violation\footnote{ Pairs of $\lambda$ discussed in 
this section enable decays $l_i \to l_j 2 \nu$ at tree-level. However, the 
experimental limits are very weak and 
thus the resulting bounds on the values of $\lambda$'s are not competitive with the ones discussed in this work.}. 
For those couplings all possible final states at
tree-level are kinematically forbidden but other decay channels are induced at 1-loop. 
 The results for all other pairs
of trilinear couplings which do open tree-level channels are given for completeness
in the appendix.

Before we present the updated bounds derived in our work, we briefly
comment on earlier results. In Ref.~\cite{deGouvea:2000cf} the old MEG
limit for Br($\mu \to e \gamma) <1.2\cdot 10^{-11}$ has been
used and the limits $|\lambda^*_{132}\cdot\lambda_{232}|<2.3\cdot10^{-4}$
and $|\lambda^*_{231} \cdot \lambda_{232}| < 8.2 \cdot 10^{-5}$ were
obtained. We have explicitly checked with our code that, using
the same experimental limit, one finds $2.1 \cdot 10^{-4}$ and $8.0
\cdot 10^{-5}$, respectively, for the same combinations of $\lambda$
couplings. This is in rather good agreement and gives an idea of the
expected theoretical uncertainty.

It has also been shown in Ref.~\cite{deGouvea:2000cf} that $\mu\to3e$
can be more constraining than $\mu \to e \gamma$. However, this result
was not based on the inclusion of the $Z^0$-penguins but 
instead on polarization effects. They set the limits $|\lambda^*_{132}
\cdot\lambda_ {232}| < 7.1\cdot 10^{-5}$ and $|\lambda^*_{231}\cdot
\lambda_{232}|<4.5\cdot 10^{-5}$. These bounds can already be reached
just by including the $Z^0$-penguins, without the necessity to consider
polarization effects. In fact,  for the spectrum of BP0 we get
\begin{equation}
\label{eq:boundsBP0meee}
|\lambda^*_{132} \cdot \lambda_{232}| < 6.8\times 10^{\text{-5}} \, \hspace{1cm} |\lambda^*_{231}\cdot \lambda_{232}| < 4.6\times 10^{\text{-5}} 
\end{equation}
\begin{table} 
\begin{tabular}{|c||c|c|c|c|} 
\hline 
Coupling & $l \to l' \gamma$ & $l \to 3 l'$ & $ \tau \to l P / \mu-e $ & $ Z \to l l' $  \\  
\hline 
 $|\lambda^*_{123} \lambda_{133}|$   & $3.2\times 10^{-2}$ & $4.8\times 10^{ {-2}}$  & $2.$                         & $2.8$  \\ 
 $|\lambda^*_{123} \lambda_{233}|$   & $2.7\times 10^{-2}$ & $5.3\times 10^{ {-2}}$  & $4.9$                        & $7.9$  \\ 
 $|\lambda^*_{132} \lambda_{232}|$   & $9.1\times 10^{-5}$ & $6.8\times 10^{ {-5}}$  & $1.5\times 10^{ {-5}}$   & $3.5$  \\ 
 $|\lambda^*_{133} \lambda_{233}|$   & $4.4\times 10^{-5}$ & $1.2\times 10^{ {-4}}$  & $2.6\times 10^{ {-5}}$   & $3.3$  \\ 
 $|\lambda^*_{231} \lambda_{232}|$   & $3.5\times 10^{-5}$ & $4.6\times 10^{ {-5}}$   & $7.7\times 10^{ {-6}}$   & $2.7$  \\ 
 $|\lambda^{',*}_{122} \lambda'_{222}|$ & $1.5\times 10^{-5}$ & $7.4\times 10^{ {-5}}$  & $1.9\times 10^{ {-5}}$   & $1.3\times 10^{-1}$  \\ 
 $|\lambda^{',*}_{123} \lambda'_{223}|$ & $1.5\times 10^{-5}$ & $7.4\times 10^{ {-5}}$  & $1.9\times 10^{ {-5}}$   & $1.3\times 10^{-1}$  \\ 
 $|\lambda^{',*}_{132} \lambda'_{232}|$ & $1.5\times 10^{-5}$ & $7.1\times 10^{ {-5}}$  & $1.9\times 10^{ {-5}}$   & $1.1\times 10^{-1}$  \\ 
 $|\lambda^{',*}_{133} \lambda'_{233}|$ & $1.5\times 10^{-5}$ & $7.1\times 10^{ {-5}}$  & $1.8\times 10^{ {-5}}$   & $1.1\times 10^{-1}$  \\ 
 $|\lambda^{',*}_{133} \lambda'_{333}|$ & $4.2\times 10^{-3}$ & $2.5\times 10^{ {-2}}$  & $5.2\times 10^{ {-2}}$   & $2.7\times 10^{-1}$  \\ 
 $|\lambda^{',*}_{233} \lambda'_{333}|$ & $4.9\times 10^{-3}$ & $2.7\times 10^{ {-2}}$  & $6.1\times 10^{ {-2}}$   & $3.0\times 10^{-1}$  \\ 
 \hline 
\end{tabular} 
\caption{New limits using our calculation evaluated at the
benchmark point BP0 on different combinations of $LLE$ and $LQD$
operators derived from low energy precision observables and the
experimental limits given in Table~\ref{tab:bounds}.  }
\label{tab:loopResults_bp0} 
\end{table} 
All the bounds evaluated using the spectrum of the benchmark
point BP0 are collected in Table~\ref{tab:loopResults_bp0}.  One can
easily see that the limits from $Z^0$ decays are very weak but all
other observables provide bounds of the same order for
most combinations of couplings. However, as already mentioned in the
introduction, both $l_i \to l_j \gamma$ and the photonic contributions
to $l_i \to 3 l_j$ and $\mu - e$ conversion in nuclei scale as
$m_{SUSY}^{-4}$ \cite{Hirsch:2012ax}.  Hence, if one only includes
these contributions all bounds are much weaker for a heavier spectrum
like in BP1 to BP3. In contrast, as shown in
\cite{Hirsch:2012ax}, $l_i \to 3 l_j$ is much less sensitive to
the SUSY scale as soon as the $Z^0$-penguins dominate: the $Z^0$
penguins are increased by a factor $m^4_{SUSY}/m^4_Z$ in comparison to
the photonic contributions. The same happens for the $Z^0$
contributions to $\mu - e$ conversion in nuclei and $\tau \to l_i P^0$
decays.
\begin{figure}[hbt]
\includegraphics[width=0.45\linewidth]{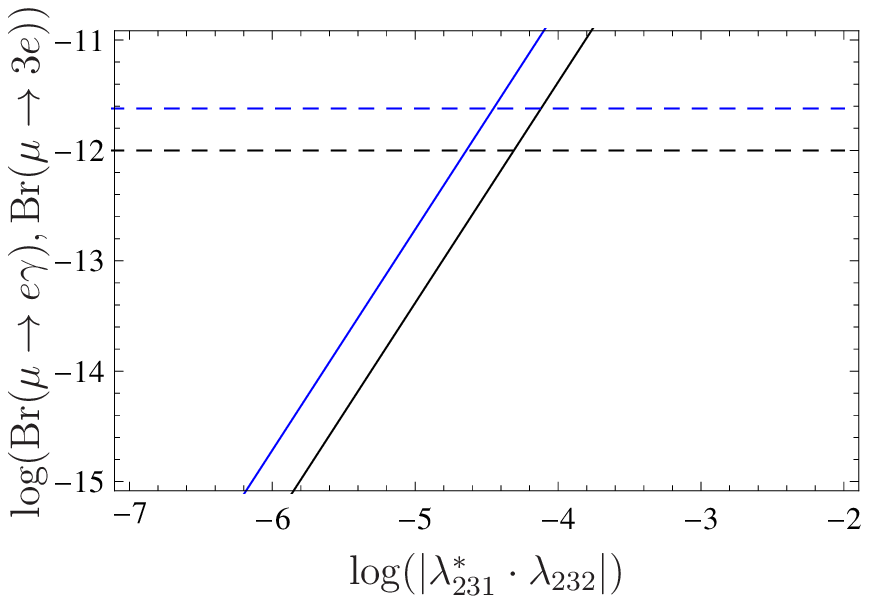} \hfill
\includegraphics[width=0.45\linewidth]{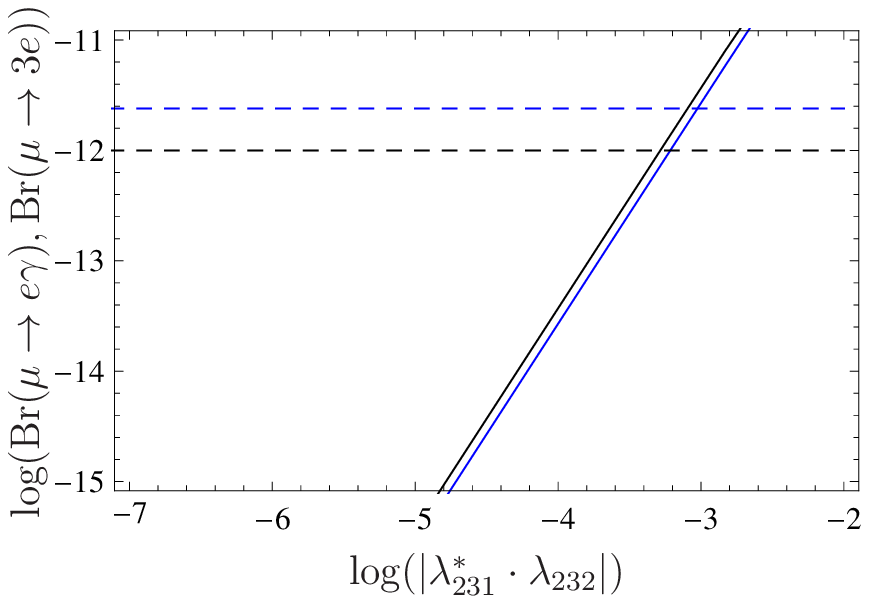}
\caption{$\text{Br}(\mu \to e \gamma)$ (blue)  and $\text{Br}(\mu\to3e)$
(black) for BP0 (left) and BP2 (right). The dashed lines show the 
current upper experimental bounds. }
\label{fig:MuE_MuEEE_bp0_bp2}
\end{figure}
To show this different behavior we depict in
Fig.~\ref{fig:MuE_MuEEE_bp0_bp2} the dependence of $\text{Br}(\mu
\to e \gamma)$ and $\text{Br}(\mu \to 3e)$ for BP0 and BP2  
on one combination of $LLE$ couplings.  While  for BP0 
$\text{Br}(\mu\to e \gamma)$ > $\text{Br}(\mu \to 3e)$ holds, the
order is changed for BP2 because $\text{Br}(\mu\to e \gamma)$ is shifted to the
right while $\text{Br}(\mu \to 3e)$ has only slightly moved.
 
Thus indeed the bounds from $l_i\rightarrow3l_j$ are less sensitive to
an increase in the SUSY mass scale. And using $\text{Br}(\mu
\to 3e)$, it is possible to derive bounds on the couplings
for the points BP1 - BP3 which are of the same order as those given in
Eq.~(\ref{eq:boundsBP0meee}) for a light SUSY spectrum. This can be
seen in Tables~\ref{tab:loopResults_bp1} to \ref{tab:loopResults_bp3},
where we give the limits of all combinations of trilinear couplings
which do not open channels for leptonic flavor violating processes at
tree-level.\footnote{With lepton flavor violating decays we refer only
to processes with three charged leptons in the final states. The
couplings will open decays $l\to l_i\nu_j\nu_k$ but those are
experimentally unconstrained.}

\begin{table} 
\begin{tabular}{|c||c|c|c|c|} 
\hline 
Coupling & $l_i \to l_j  \gamma$ & $l_i \to 3 l_j$ & $ \tau \to l_i P / \mu-e $ & $ Z^0 \to l_i l_j $  \\  
\hline 
 $|\lambda^*_{123} \lambda_{133}|$ & $5.5\times 10^{-1}$   & $4.8\times 10^{ {-1}}$  & $3.4\times 10^1$            & $4.5$  \\ 
 $|\lambda^*_{123} \lambda_{233}|$ & $4.8\times 10^{-1}$   & $5.4\times 10^{ {-1}}$  & $5.3$                       & $1.3\times 10^1$  \\ 
 $|\lambda^*_{132} \lambda_{232}|$ & $2.3\times 10^{-3}$   & $8.2\times 10^{ {-4}}$  & $1.6\times 10^{ {-4}}$  & $5.8$  \\ 
 $|\lambda^*_{133} \lambda_{233}|$ & $5.6\times 10^{-4}$   & $1.1\times 10^{ {-3}}$  & $2.2\times 10^{ {-4}}$  & $5.4$  \\ 
 $|\lambda^*_{231} \lambda_{232}|$ & $3.8\times 10^{-4}$   & $4.1\times 10^{ {-4}}$  & $1.2\times 10^{ {-4}}$  & $9.7$  \\ 
 $|\lambda^{',*}_{122} \lambda'_{222}|$ & $1.2\times 10^{-4}$ & $5.0\times 10^{ {-5}}$  & $1.0\times 10^{ {-5}}$   & $1.8$  \\ 
 $|\lambda^{',*}_{123} \lambda'_{223}|$ & $1.2\times 10^{-4}$ & $5.0\times 10^{ {-5}}$  & $1.0\times 10^{ {-5}}$   & $1.8$  \\ 
 $|\lambda^{',*}_{132} \lambda'_{232}|$ & $1.3\times 10^{-4}$ & $5.3\times 10^{ {-5}}$  & $1.1\times 10^{ {-5}}$  & $8.1\times 10^{-1}$  \\ 
 $|\lambda^{',*}_{133} \lambda'_{233}|$ & $1.3\times 10^{-4}$ & $5.3\times 10^{ {-5}}$  & $1.1\times 10^{ {-5}}$  & $8.1\times 10^{-1}$  \\ 
 $|\lambda^{',*}_{133} \lambda'_{333}|$ & $3.3\times 10^{-2}$ & $2.1\times 10^{ {-2}}$  & $3.7\times 10^{ {-2}}$  & $1.9$  \\ 
 $|\lambda^{',*}_{233} \lambda'_{333}|$ & $3.8\times 10^{-2}$ & $1.8\times 10^{ {-2}}$  & $4.3\times 10^{ {-2}}$  & $2.2$  \\ 
 \hline 
\end{tabular} 
\caption{Limits for the benchmark point BP1 on different combinations of 
$LLE$ and $LQD$ operators derived from low energy precision observables and 
the experimental limits given in Table~\ref{tab:bounds}.}
\label{tab:loopResults_bp1} 
\end{table} 
\begin{table} 
\begin{tabular}{|c||c|c|c|c|} 
\hline 
Coupling & $l_i \to l_j  \gamma$ & $l_i \to 3 l_j$ & $ \tau \to l_i P / \mu-e $ & $ Z^0 \to l_i l_j $  \\  
\hline 
 $|\lambda^*_{123} \lambda_{133}|$ & $1.8\times 10^1$      & $1.2$                       & $8.3\times 10^1$            & $1.4\times 10^1$  \\ 
 $|\lambda^*_{123} \lambda_{233}|$ & $1.3\times 10^1$      & $1.4$                       & $5.9$                       & $4.\times 10^1$  \\ 
 $|\lambda^*_{132} \lambda_{232}|$ & $2.4\times 10^{-1}$   & $2.2\times 10^{ {-3}}$  & $4.2\times 10^{ {-4}}$  & $1.7\times 10^1$  \\ 
 $|\lambda^*_{133} \lambda_{233}|$ & $1.7\times 10^{-3}$   & $3.0\times 10^{ {-3}}$  & $6.1\times 10^{ {-4}}$  & $1.7\times 10^1$  \\ 
 $|\lambda^*_{231} \lambda_{232}|$ & $9.5\times 10^{-4}$   & $5.2\times 10^{ {-4}}$  & $2.4\times 10^{ {-4}}$  & $2.3\times 10^1$  \\ 
 $|\lambda^{',*}_{122} \lambda'_{222}|$ & $4.5\times 10^{-4}$ & $4.3\times 10^{ {-5}}$  & $8.8\times 10^{ {-6}}$  & $7.5\times 10^{-1}$  \\ 
 $|\lambda^{',*}_{123} \lambda'_{223}|$ & $4.6\times 10^{-4}$ & $4.3\times 10^{ {-5}}$  & $9.0\times 10^{ {-6}}$  & $7.5\times 10^{-1}$  \\ 
 $|\lambda^{',*}_{132} \lambda'_{232}|$ & $4.9\times 10^{-4}$ & $4.5\times 10^{ {-5}}$  & $9.3\times 10^{ {-6}}$  & $1.4$  \\ 
 $|\lambda^{',*}_{133} \lambda'_{233}|$ & $4.9\times 10^{-4}$ & $4.5\times 10^{ {-5}}$  & $9.3\times 10^{ {-6}}$  & $1.4$  \\ 
 $|\lambda^{',*}_{133} \lambda'_{333}|$ & $1.3\times 10^{-1}$ & $1.8\times 10^{ {-2}}$  & $3.1\times 10^{ {-2}}$  & $3.3$  \\ 
 $|\lambda^{',*}_{233} \lambda'_{333}|$ & $1.5\times 10^{-1}$ & $1.6\times 10^{ {-2}}$  & $3.6\times 10^{ {-2}}$  & $3.6$  \\ 
 \hline 
\end{tabular} 
\caption{Limits for BP2 on different combinations of $LLE$ and $LQD$ operators derived from low energy precision observables and 
the experimental limits given in Table~\ref{tab:bounds}.} 
\label{tab:loopResults_bp2} 
\end{table} 
\begin{table} 
\begin{tabular}{|c||c|c|c|c|} 
\hline 
Coupling & $l_i \to l_j  \gamma$ & $l_i \to 3 l_j$ & $ \tau \to l_i P / \mu-e $ & $ Z^0 \to l_i l_j $  \\  
\hline 
 $|\lambda^*_{123} \lambda_{133}|$ & $1.2\times 10^1$             & $2.4$                       & $6.9$                        & $2.\times 10^1$  \\ 
 $|\lambda^*_{123} \lambda_{233}|$ & $1.2\times 10^1$             & $2.8$                       & $2.1\times 10^{ {-1}}$   & $5.7\times 10^1$  \\ 
 $|\lambda^*_{132} \lambda_{232}|$ & $3.4\times 10^{ {-3}}$   & $3.3\times 10^{ {-3}}$  & $6.5\times 10^{ {-4}}$   & $6.1\times 10^1$  \\ 
 $|\lambda^*_{133} \lambda_{233}|$ & $1.9\times 10^{ {-3}}$   & $4.5\times 10^{ {-3}}$  & $9.2\times 10^{ {-4}}$   & $2.8\times 10^1$  \\ 
 $|\lambda^*_{231} \lambda_{232}|$ & $3.1\times 10^{ {-3}}$   & $4.7\times 10^{ {-4}}$  & $1.3\times 10^{ {-4}}$   & $3.6\times 10^1$  \\ 
 $|\lambda^{',*}_{122} \lambda'_{222}|$ & $3.\times 10^{ {-4}}$  & $4.3\times 10^{ {-5}}$  & $9.0\times 10^{ {-6}}$    & $8.9\times 10^{ {-1}}$  \\ 
 $|\lambda^{',*}_{123} \lambda'_{223}|$ & $3.3\times 10^{ {-4}}$ & $4.4\times 10^{ {-5}}$  & $9.0\times 10^{ {-6}}$    & $8.9\times 10^{ {-1}}$  \\ 
 $|\lambda^{',*}_{132} \lambda'_{232}|$ & $3.4\times 10^{ {-4}}$ & $4.7\times 10^{ {-5}}$  & $9.1\times 10^{ {-6}}$  & $6.7$  \\ 
 $|\lambda^{',*}_{133} \lambda'_{233}|$ & $3.8\times 10^{ {-4}}$ & $4.7\times 10^{ {-5}}$  & $9.7\times 10^{ {-6}}$   & $8.6$  \\ 
 $|\lambda^{',*}_{133} \lambda'_{333}|$ & $8.7\times 10^{ {-2}}$ & $1.8\times 10^{ {-2}}$  & $2.2\times 10^{ {-2}}$   & $2.1\times 10^1$  \\ 
 $|\lambda^{',*}_{233} \lambda'_{333}|$ & $9.8\times 10^{ {-2}}$ & $1.6\times 10^{ {-2}}$  & $3.8\times 10^{ {-2}}$   & $2.3\times 10^1$  \\ 
 \hline 
\end{tabular} 
\caption{Limits for BP3 on different combinations of $LLE$ and $LQD$ operators derived from low energy precision observables and 
the experimental limits given in Table~\ref{tab:bounds}.} 
\label{tab:loopResults_bp3} 
\end{table} 
 Thus as discussed above, the bounds coming from observables
which involve $Z^0$ penguin diagrams depend only very 
 mildly on the SUSY point.  In fact, some bounds even
get improved slightly with a heavier mass spectrum. This is more
pronounced in case of $LQD$ couplings.  In particular, BP2 and BP3 are
a bit more restrictive than BP1 and BP0.  The reason for this can be
found in the wave function contributions to the $Z^0$ penguins
involving the loop function $B_1$~\cite{Arganda:2005ji}
\begin{equation}
 B_1(m_q^2,m_{\tilde{q}}^2) = - \frac{1}{2} + \frac{1}{2}\text{log}
(m^2_{\tilde{q}}) - \frac{m^2_q - m^2_{\tilde{q}} + 2 m_1^2 \text{log}
(\frac{m^2_{\tilde{q}}}{m_q^2})}{4(m_q^2 - m^2_{\tilde{q}})^2}
\end{equation}
with quark mass $m_q$ and squark mass $m_{\tilde{q}}$. Hence, these 
contributions grow logarithmically with the scalar masses in the loop. 

The combinations $|\lambda^*_{123}\lambda_{233}|$, $|\lambda^*_{123}\lambda_
{133}|$, $|\lambda^{',*}_{133} \lambda'_{333}|$ and $|\lambda^{',*}_{233}\lambda'_
{333}|$ are less  constrained than the other $|\lambda^*\lambda|$ or
$|\lambda^{',*} \lambda' |$ combinations because they induce $\tau$ decays
while all other combinations contribute to $\mu$ decays. Nevertheless,
these combinations show in general the same qualitative behavior when
the different  benchmark points are compared.

A final comment about the lepton flavor violating three-body decays:
while the derived bounds on $|\lambda^*_{132} \lambda_{232}|$ and
$|\lambda^*_{133} \lambda_{233}|$ are of the same size, $|\lambda^*_{231}
\lambda_{232}|$ is always a bit more  constrained. The difference
between these contributions is that for the first two combinations the
charged lepton can be right-handed while for the third case the lepton
has to be left-handed and has therefore a larger coupling to the $Z^0$
boson.
\begin{figure}[hbt]
\includegraphics[width=0.45\linewidth]{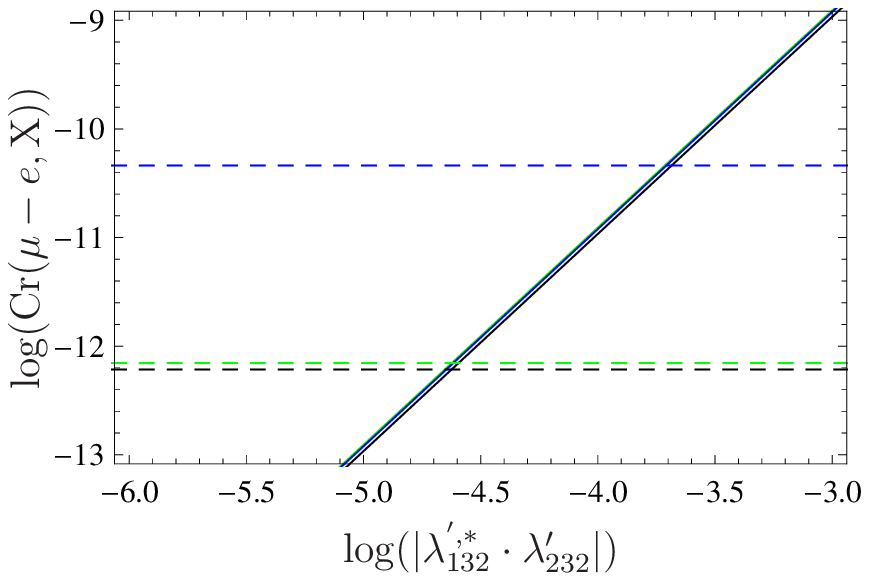} \hfill
\includegraphics[width=0.45\linewidth]{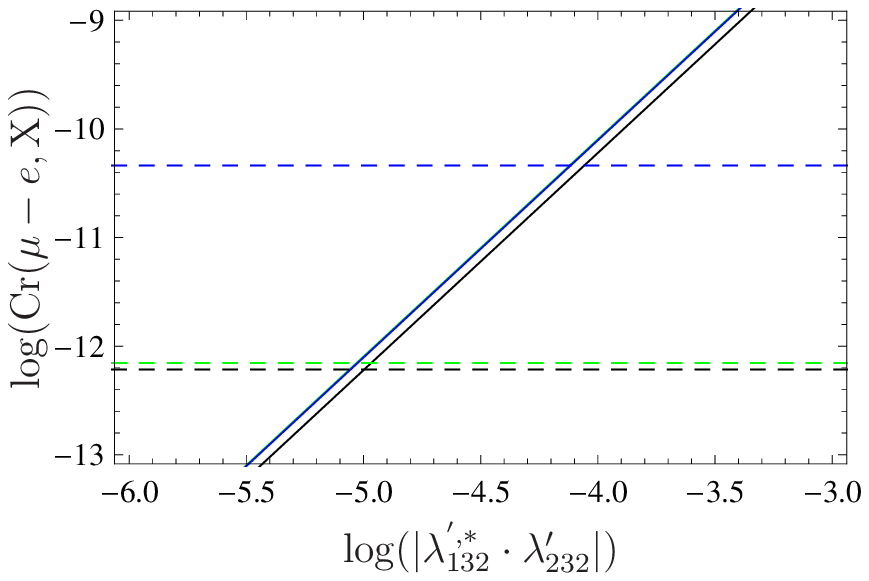}
\caption{$\text{Cr}(\mu - e, \text{Au})$ (green), $\text{Cr}(\mu - e, \text{Ti})$ (black)  
and $\text{Cr}(\mu - e, \text{Pb})$ (blue) for  BP0 (left) and BP2 (right)  
as function of $\text{log}(|\lambda_{132}^{',*} \lambda'_{232}|)$. 
The dashed lines show the current  upper experimental  bounds. }
\label{fig:mu-e_bp0_bp2}
\end{figure}

$\mu - e$ conversion in nuclei in the context of trilinear $R$-parity
violation was also studied in Ref. \cite{deGouvea:2000cf}. The limit
obtained for instance for $|\lambda^*_{132} \lambda_{232}|$ was $1.3\cdot
10^{-5}$. This bound is based on the same experimental limit of
$\text{Cr}(\mu -e, \text{Ti})$ given in Table~\ref{tab:bounds} for
which we get nearly the same value as for gold nuclei, namely
$|\lambda^*_{132}\lambda_{232}| <  1.5\cdot 10^{-5}$.  

In general, in most cases $\mu - e$ conversion in nuclei or $\tau \to
l_i P^0$ can be used to derive even stricter limits than those given
by the three-body decays. The main reason for this is the very good
experimental limit due to $\mu - e$ conversion in gold and, of course,
the same small dependence on the SUSY masses due to unsuppressed
$Z^0$-penguins. This can be seen in Fig.~\ref{fig:mu-e_bp0_bp2}. The
main points of the discussion about the limits given by loop induced
three-body decays apply also here.  However, there is one additional,
interesting observation: $\mu-e$ conversion in nuclei leads in the
case of $LQD$ couplings to a constraint for BP1 which is better than
the one for BP0 by a factor of 2. This effect is larger than in the
case of $l_i\to3l_j$ decays and not only caused by the logarithmic
growth of the wave function contributions.  The main reason for the
difference in the bounds comes from the photon contributions to $\mu -
e$ conversion which are, for BP0, of the same size as the $Z^0$
penguins. This leads to a negative interference reducing the severity
of the limits. The very heavy squarks in the case of BP2 and BP3 are
reflected by the very good limits for $\mu - e$ conversion for $LQD$
couplings while the bounds from $LLE$  are better for BP1
than for BP2. If the future plans to reach a sensitivity for the $\mu - e$
conversion rate in Titanium of  $10^{-18}$ \cite{Carey:2008zz} succeed,
and no anomaly is observed, the corresponding limits are expected to
improve by three orders of magnitude, e.g.  BP2 would set a limit for
$|\lambda^*_{231}\lambda_{232}|$ of   $4.3\cdot 10^{-7}$.

\begin{figure}[hbt]
\includegraphics[width=0.6\linewidth]{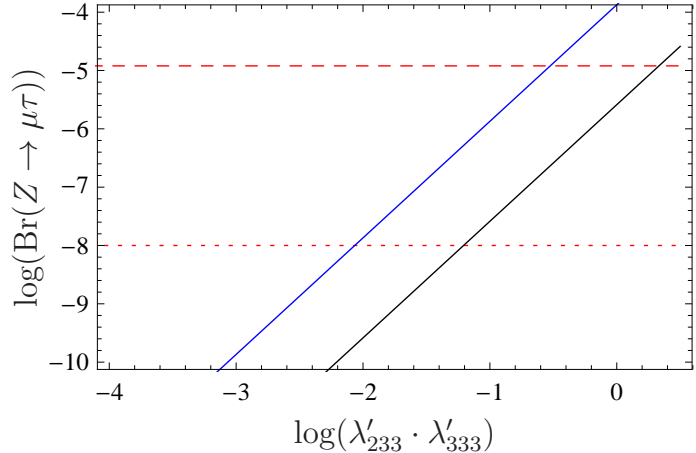} 
\caption{$\text{Br}(Z^0 \to \mu \tau)$ for BP0 (blue) and BP1 (black).
  The red dashed line corresponds to the current experimental LEP
  limit of $1.2\cdot 10^{-5}$ \cite{Nakamura:2010zzi}, the red
  dot-dashed line shows the limit of $1.0\cdot 10^{-8}$ which might be
  reached by Giga-Z \cite{Illana:1999ww}. }
\label{fig:Zmutau_bp0_bp1}
\end{figure}

Finally, we comment on rare $Z^0$ decays. The flavor violating decays
of the $Z^0$ gauge boson do not set new constraints on the
parameters. In fact, for many combinations of couplings the resulting
limits could only be estimated by extrapolation since they lie already
in the non-perturbative regime. Only when heavy quarks are present in
the loop  could the $Z^0$ decays be of some
relevance.  Using the expected experimental limits of Giga-Z
\cite{Illana:1999ww} the $Z^0$ decays into $\mu \tau$ might reach the
importance of the other observables.  An estimate of the potential
improvement on the bounds is shown in Fig.~\ref{fig:Zmutau_bp0_bp1}.
We considered a future limit of $1.0\cdot 10^{-8}$ for $\text{Br}(Z^0
\to \mu \tau)$ and found a limit of $O(10^{-2})$ on the  product
of the couplings.  However, in case of lepton flavor violation in the
$\mu -e$ sector, the $Z^0$ decays will never reach the current
sensitivity of $l_i \to 3 l_j$ or $\mu - e$ conversion in nuclei. To
get a comparable limit, for instance for $|\lambda^*_{132} \lambda_{232}|$
in case of BP3 of $O(10^{-5})$, the limit of $\text{Br}(Z^0 \to \mu
e)$ should be improved to $O(10^{-19})$ which is far  beyond the
reach of the ILC with Giga-Z.

\section{Conclusion}
We have considered in this paper the bounds on different combinations
of $LLE$ and $LQD$ operators in case of trilinear $R$-parity violation
obtained from the experimental limits on different low energy
observables. We have taken into account the 1-loop induced flavor
violating decays $l_i \to l_j \gamma$, $l_i \to 3 l_j$, $\tau \to l_i
P^0$ and $Z^0 \to l_i l_j$ as well as $\mu-e$ conversion in nuclei. It
turns out that the $Z^0$ penguins dominate in most parts of parameter
space, and especially for heavy SUSY spectra, the amplitudes for $l_i
\to 3 l_j$, $\tau \to l_i P^0$ and $\mu - e$ conversion.  Therefore,
the limits on combinations of $\lambda$ and $\lambda'$ couplings given
by these observables change only slightly between the different
benchmark points. Taking into account the most stringent observables,
$\mu - e$ conversion in nuclei and $\tau \to l_i P^0$ decays, one
finds for heavy SUSY scenarios improvements of several orders of
magnitude with respect to the bounds already present in the
literature.

\section*{Acknowledgements}

We thank Martin Hirsch and Werner Porod for fruitful discussions.
A.V. acknowledges support from the ANR project CPV-LFV-LHC
{NT09-508531}.

\begin{appendix}
\section{Tree-level induced decays $l_i \to 3 l_j$ and $l_i \to l_j l_k l_k$
in R-parity Violation}
\label{sec:treelevel}
\begin{figure}
\centering
\includegraphics[width=0.45\linewidth]{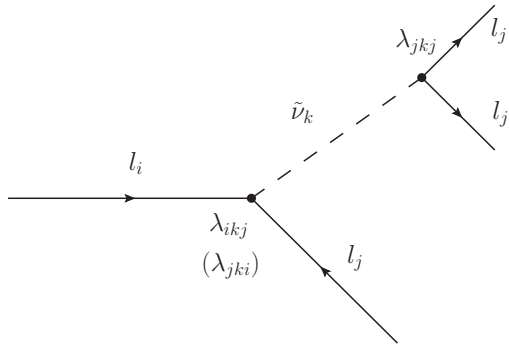}
\caption{Tree-level induced $l_i \to 3 l_j$ decays. As shown in 
brackets, there are two possible combinations of $\lambda$ couplings:
$\lambda_{jki} \lambda_{iki}$ and $\lambda_{ikj}
\lambda_{iki}$. Moreover, we remind the reader that the $\lambda$
couplings are antisymmetric in the first two indices.}
\label{tree-diag-1}
\end{figure}

\begin{figure}
\centering
\subfigure[]{
\includegraphics[width=0.45\linewidth]{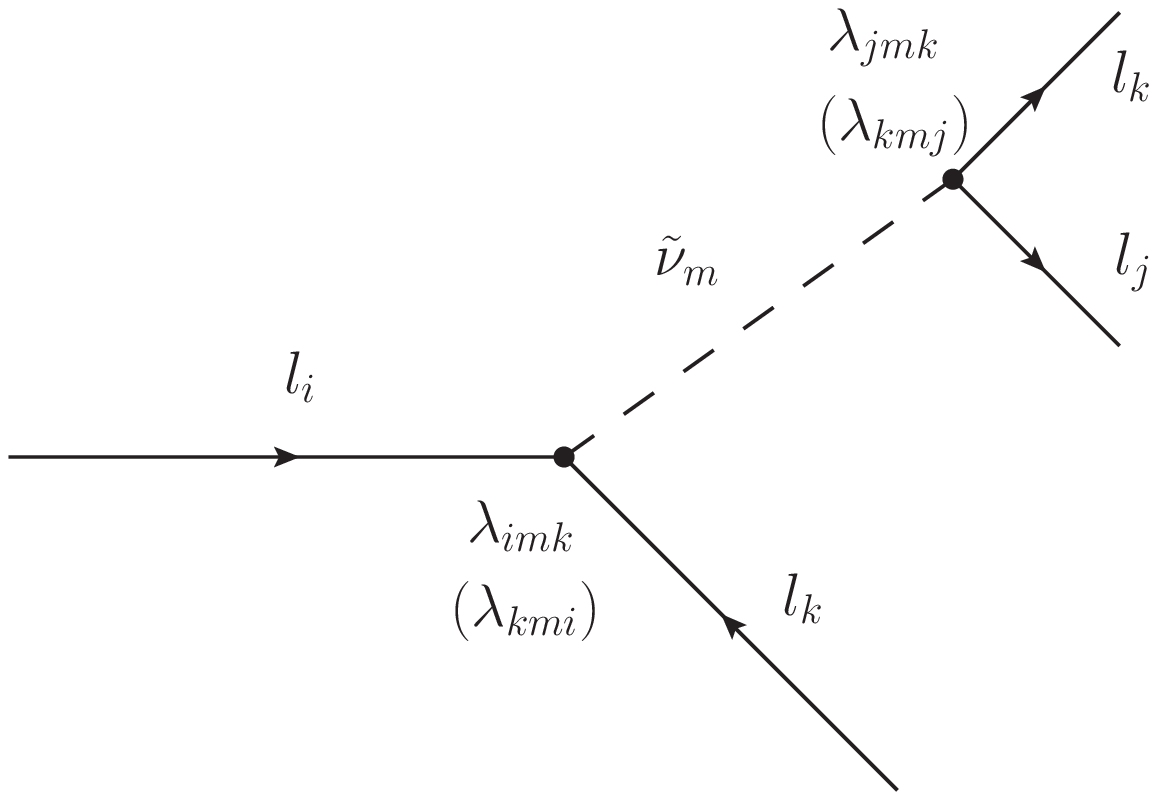}
\label{fig:tree-2-1}
}
\subfigure[]{
\includegraphics[width=0.45\linewidth]{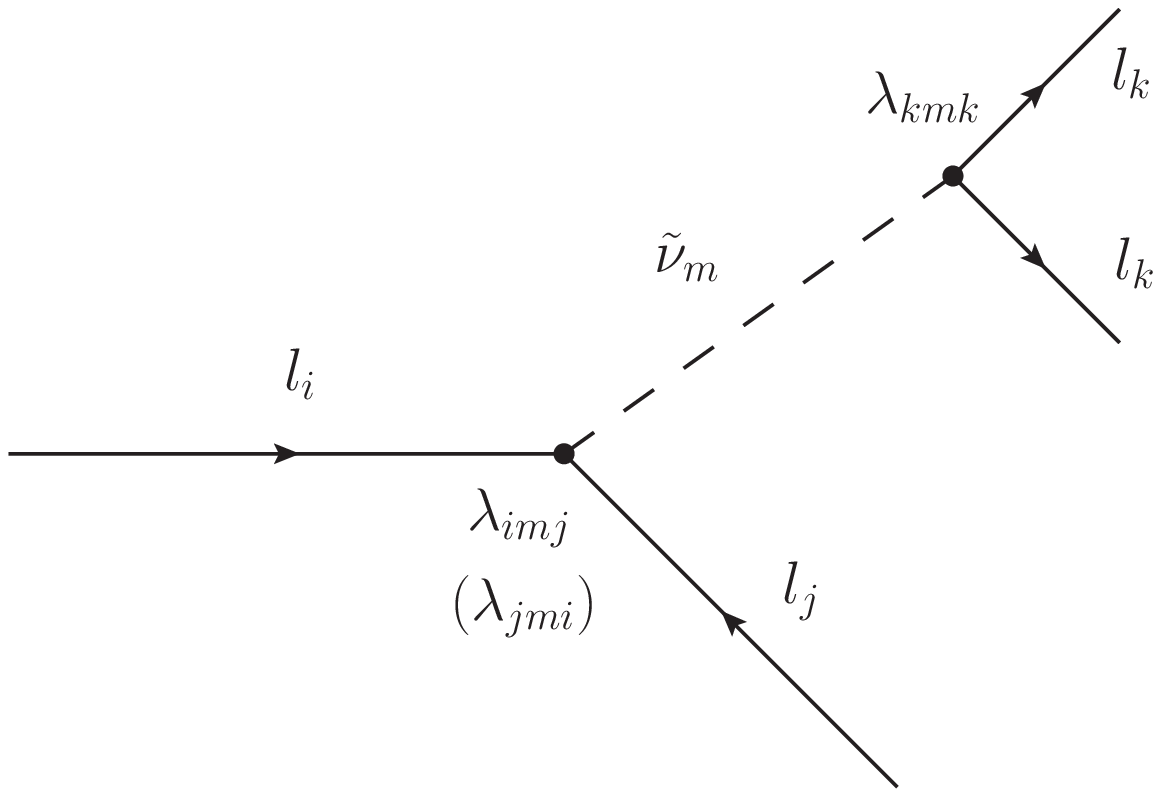}
\label{fig:tree-2-2}
}
\caption{Tree-level induced $l_i \to l_j l_k l_k$ decays ($j \ne k$). The different indices combinations are shown in brackets. Case \subref{fig:tree-2-1}: $\lambda_{jmk} \lambda_{imk}$, $\lambda_{jmk} \lambda_{kmi}$, $\lambda_{kmj} \lambda_{imk}$ and $\lambda_{kmj} \lambda_{kmi}$. Case \subref{fig:tree-2-2}: $\lambda_{jmi} \lambda_{kmk}$ and $\lambda_{imj} \lambda_{kmk}$. Moreover, we remind the reader that the $\lambda$ couplings are antisymmetric in the first two indices.}
\label{tree-diag-2}
\end{figure}

As already mentioned, specific combinations of $\lambda$ and $\lambda'$ open 
lepton flavor violating decay channels already at tree-level. In this context,
both $l_i \to 3 l_j$ and $l_i \to l_j l_k l_k$ have already been
studied in detail in the literature, see for example
Refs.~\cite{Choudhury:1996ia,deGouvea:2000cf}. Since several sneutrino
mediated diagrams exist, see Fig.~\ref{tree-diag-1} (for $l_i \to 3
l_j$) and Fig.~\ref{tree-diag-2} (for $l_i \to l_j l_k l_k$, with $j
\ne k$), quite a few combinations of $\lambda \lambda$ parameters can
be constrained.

One can compute the corresponding branching ratios by means of the
effective 4-fermion operator obtained after integrating out the
sneutrino \cite{Choudhury:1996ia}. This possibility is perfectly valid
due to the large hierarchy between the masses of the charged leptons
and the mass of the sneutrino. However, we have taken a different
approach, based on the exact computation of the tree-level diagrams,
with full 3-body phase space evaluation and including the widths of
the sneutrinos. 

In addition to the bounds given in Table~\ref{tab:bounds}, we use for
the tree-level decays observables with two different leptons in the
final state.  The  experimental upper bounds on the respective
branching ratios are
\cite{Nakamura:2010zzi}
\begin{align}
 \label{eq:limit3}
& \tau^- \to \mu^- e^+ e^-: \, \, 1.8 \cdot 10^{-8}, \hspace{0.5cm} 
\tau^- \to \mu^+ e^- e^-: \, \, 1.5 \cdot 10^{-8} \\ 
& \label{eq:limit4} 
 \tau^- \to e^- \mu^+ \mu^-: \, \, 2.7 \cdot 10^{-8} , \hspace{0.5cm}  
\tau^- \to e^+ \mu^- \mu^-: \, \, 2.7 \cdot 10^{-8} 
\end{align}
The bounds obtained by these observables are presented in
Table~\ref{tab:limits_tree}.  It can be seen that the bounds for
couplings which open the $\mu \to 3e$ decay mode are in agreement with
\cite{deGouvea:2000cf} for BP0. All other bounds are also compatible
if one considers the usual $\sim m_{SUSY}^{-4}$ scaling and in general
the limits of couplings which are only sensitive to $l_i \to l_j l_k l_l$ 
 are much weaker than those for couplings which enable also $l_i \to 3 l_k$. 
In addition, it is interesting to see that the bounds on $R$pV couplings at tree-level 
in general are not much better than those derived at 1-loop. The reason is, of
course, the different scaling of the $Z^0$-penguin.

\begin{table}[hbt]
\begin{tabular}{|c|c|c|c|c|} 
\hline 
 & BP0 &  BP1 &  BP2 &  BP3 \\ 
\hline 
\hline 
$|\lambda^*_{1 2 1} \lambda_{1 2 2}|$ & $5.1\times 10^{\text{-7}}$  & $6.2\times 10^{\text{-6}}$  & $1.9\times 10^{\text{-5}}$  & $4.0\times 10^{\text{-5}}$  \\ 
$|\lambda^*_{1 2 1} \lambda_{1 2 3}|$ & $2.2\times 10^{\text{-4}}$  & $2.6\times 10^{\text{-3}}$  & $8.4\times 10^{\text{-3}}$  & $1.7\times 10^{\text{-2}}$  \\ 
$|\lambda^*_{1 2 1} \lambda_{1 3 1}|$ & $1.7\times 10^{\text{-2}}$  & $2.0\times 10^{\text{-1}}$  & $2.3\times 10^{\text{-1}}$  & $1.2$  \\ 
$|\lambda^*_{1 2 1} \lambda_{1 3 2}|$ & $1.9\times 10^{\text{-2}}$  & $2.3\times 10^{\text{-1}}$  & $1.5\times 10^{\text{-1}}$  & $1.4$  \\ 
$|\lambda^*_{1 2 1} \lambda_{2 3 1}|$ & $2.2\times 10^{\text{-4}}$  & $2.6\times 10^{\text{-3}}$  & $8.4\times 10^{\text{-3}}$  & $1.7\times 10^{\text{-2}}$  \\ 
$|\lambda^*_{1 2 1} \lambda_{2 3 2}|$ & $1.7\times 10^{\text{-2}}$  & $2.0\times 10^{\text{-1}}$  & $6.0\times 10^{\text{-1}}$  & $1.2$  \\ 
$|\lambda^*_{1 2 2} \lambda_{1 2 3}|$ & $2.0\times 10^{\text{-4}}$  & $2.4\times 10^{\text{-3}}$  & $3.5\times 10^{\text{-3}}$  & $1.5\times 10^{\text{-2}}$  \\ 
$|\lambda^*_{1 2 2} \lambda_{1 3 1}|$ & $1.9\times 10^{\text{-2}}$  & $2.3\times 10^{\text{-1}}$  & $2.6\times 10^{\text{-1}}$  & $1.4$  \\ 
$|\lambda^*_{1 2 2} \lambda_{1 3 2}|$ & $2.0\times 10^{\text{-4}}$  & $2.4\times 10^{\text{-3}}$  & $3.5\times 10^{\text{-3}}$  & $1.6\times 10^{\text{-2}}$  \\ 
$|\lambda^*_{1 2 2} \lambda_{2 3 1}|$ & $1.7\times 10^{\text{-2}}$  & $2.0\times 10^{\text{-1}}$  & $6.0\times 10^{\text{-1}}$  & $1.2$  \\ 
$|\lambda^*_{1 2 2} \lambda_{2 3 2}|$ & $1.9\times 10^{\text{-2}}$  & $2.3\times 10^{\text{-1}}$  & $6.7\times 10^{\text{-1}}$  & $1.4$  \\ 
$|\lambda^*_{1 3 1} \lambda_{1 3 2}|$ & $4.9\times 10^{\text{-7}}$  & $6.1\times 10^{\text{-6}}$  & $6.2\times 10^{\text{-7}}$  & $3.1\times 10^{\text{-5}}$  \\ 
$|\lambda^*_{1 3 1} \lambda_{1 3 3}|$ & $2.2\times 10^{\text{-4}}$  & $2.6\times 10^{\text{-3}}$  & $2.5\times 10^{\text{-4}}$  & $1.4\times 10^{\text{-2}}$  \\ 
$|\lambda^*_{1 3 1} \lambda_{2 3 1}|$ & $4.9\times 10^{\text{-7}}$  & $6.1\times 10^{\text{-6}}$  & $6.2\times 10^{\text{-7}}$  & $3.1\times 10^{\text{-5}}$  \\ 
$|\lambda^*_{1 3 1} \lambda_{2 3 3}|$ & $1.7\times 10^{\text{-2}}$  & $2.0\times 10^{\text{-1}}$  & $2.0\times 10^{\text{-2}}$  & $1.0$  \\ 
$|\lambda^*_{1 3 2} \lambda_{1 3 3}|$ & $3.5\times 10^{\text{-3}}$  & $6.8\times 10^{\text{-3}}$  & $2.0\times 10^{\text{-2}}$  & $1.0$  \\ 
$|\lambda^*_{1 3 2} \lambda_{2 3 3}|$ & $1.9\times 10^{\text{-2}}$  & $2.3\times 10^{\text{-1}}$  & $2.3\times 10^{\text{-2}}$  & $1.1$  \\ 
$|\lambda^*_{1 3 3} \lambda_{2 3 1}|$ & $1.7\times 10^{\text{-2}}$  & $2.0\times 10^{\text{-1}}$  & $2.0\times 10^{\text{-2}}$  & $1.0$  \\ 
$|\lambda^*_{1 3 3} \lambda_{2 3 2}|$ & $1.9\times 10^{\text{-2}}$  & $2.2\times 10^{\text{-1}}$  & $2.3\times 10^{\text{-2}}$  & $1.1$  \\ 
$|\lambda^*_{2 3 1} \lambda_{2 3 3}|$ & $1.9\times 10^{\text{-2}}$  & $4.3\times 10^{\text{-2}}$  & $2.3\times 10^{\text{-2}}$  & $1.1$  \\ 
$|\lambda^*_{2 3 2} \lambda_{2 3 3}|$ & $2.0\times 10^{\text{-4}}$  & $2.4\times 10^{\text{-3}}$  & $2.3\times 10^{\text{-4}}$  & $1.3\times 10^{\text{-2}}$  \\ 
\hline 
\end{tabular} 
\caption{Bounds on combinations of $LLE$ couplings from the LFV decays $l_i \to 3 l_j$ and $l_i \to l_j l_k l_k$ induced at tree-level.}
\label{tab:limits_tree}
\end{table}

\end{appendix}

\end{document}